\documentclass[aip,jap,10pt,twocolumn,superscriptaddress]{revtex4-1}
\usepackage{graphicx}
\usepackage{amsmath,amssymb}
\usepackage{txfonts}
\usepackage{bm}
\usepackage[usenames,dvipsnames]{xcolor}

\begin{document}

\title{Magnetoelectric domain wall dynamics and its implications for magnetoelectric memory}

\author{K. D. Belashchenko}
\affiliation{Department of Physics and Astronomy and Nebraska Center for Materials and Nanoscience, University of Nebraska-Lincoln, Lincoln, Nebraska 68588, USA}

\author{O. Tchernyshyov}
\affiliation{Department of Physics and Astronomy, Johns Hopkins University,  Baltimore, Maryland 21218, USA}

\author{Alexey A. Kovalev}
\affiliation{Department of Physics and Astronomy and Nebraska Center for Materials and Nanoscience, University of Nebraska-Lincoln, Lincoln, Nebraska 68588, USA}

\author{O. A. Tretiakov}
\affiliation{Institute for Materials Research, Tohoku University, Sendai 980-8577, Japan}
\affiliation{School of Natural Sciences, Far Eastern Federal University, Vladivostok 690950, Russia}

\date{\today}

\begin{abstract}
Domain wall dynamics in a magnetoelectric antiferromagnet is analyzed, and its implications for magnetoelectric memory applications are discussed. Cr$_2$O$_3$ is used in the estimates of the materials parameters. It is found that the domain wall mobility has a maximum as a function of the electric field due to the gyrotropic coupling induced by it. In Cr$_2$O$_3$ the maximal mobility of 0.1 m/(s$\times$Oe) is reached at $E\approx0.06$ V/nm. Fields of this order may be too weak to overcome the intrinsic depinning field, which is estimated for B-doped Cr$_2$O$_3$. These major drawbacks for device implementation can be overcome by applying a small in-plane shear strain, which blocks the domain wall precession. Domain wall mobility of about 0.7 m/(s$\times$Oe) can then be achieved at $E=0.2$ V/nm. A split-gate scheme is proposed for the domain-wall controlled bit element; its extension to multiple-gate linear arrays can offer advantages in memory density, programmability, and logic functionality.
\end{abstract}

\maketitle

Encoding and manipulation of information by the antiferromagnetic (AFM) order parameter has recently attracted considerable attention,\cite{MacDonald,Gomonay,Jungwirth,Marti} and current-induced switching of a metallic antiferromagnet has been demonstrated.\cite{Wadley} Device concepts utilizing a magnetoelectric antiferromagnet (MEAF) as the active element are also being actively pursued for applications in nonvolatile memory and logic.\cite{BD,He,Manipatruni} The fundamental principle of operation involves the reversal of the AFM order parameter in the MEAF by applied voltage in the presence of an external magnetic field, which is accompanied by the reversal of the boundary magnetization of the MEAF.\cite{He,Andreev,Belashchenko} Little is known, however, about the fundamental limitations of this approach. Here we discuss the switching mechanisms, describe the dynamics of a moving domain wall, estimate the relevant metrics, and propose the scheme of a memory bit.

We consider the usual case of a collinear MEAF, such as Cr$_2$O$_3$, with two macroscopically inequivalent AFM domains, mapped one onto the other by time reversal. The driving force for the switching of such MEAF is the difference $F=2\mathbf{E}\hat\alpha\mathbf{H}$ in the free energy densities of the two AFM domains, where $\hat\alpha$ is the magnetoelectric tensor.\cite{LL} Thermally activated single-domain switching involves a severe tradeoff between thermal stability and switching time --- a long-standing problem in magnetic recording technology.\cite{HAMR} In order to significantly reduce the activation barrier for single-domain switching, the applied fields should satisfy $\alpha EH\sim K$, where $K$ is the magnetocrystalline anisotropy constant. In Cr$_2$O$_3$, where $\alpha\lesssim 10^{-4}$ (Gaussian units) and $K\approx 2\times 10^5$ erg/cm$^3$,\cite{Borovik,Foner} this condition requires $EH\sim10^{11}$ Oe$\times$V/cm. Since fields of this magnitude are undesirable for device applications, we are led to consider inhomogeneous switching, which involves nucleation of reverse domains and domain wall motion. The switching time is determined by the slower of these two mechanisms. Nucleation is a relatively slow thermally-activated process, which can be avoided by device engineering, as discussed below. The switching time is then limited by the domain wall motion driven by the magnetoelectric pressure $F$.

The magnetic dynamics in an AFM is qualitatively different from a ferromagnet (FM).\cite{Bar,Haldane,Kim} If the magnetostatic interaction is neglected, a domain wall in an ideal FM with no damping does not move at all, but rather precesses in the applied magnetic field. The FM domain wall velocity $v$ in this case is proportional to the small Gilbert damping parameter $\alpha_0$. The magnetostatic interaction lifts the degeneracy of the Bloch and N\'eel configurations and blocks the precession, making $v\propto\alpha_0^{-1}$ as long as $v$ does not exceed the Walker breakdown velocity $v_W$.\cite{Walker} In contrast, in AFM the Gilbert damping limits the terminal velocity of the wall. Here we are interested in the dynamics of a domain wall in a MEAF, such as Cr$_2$O$_3$, which is driven by the application of the electric and magnetic fields. In a finite electric field, a MEAF turns into a nearly-compensated ferrimagnet. As we will see below, the existence of a small magnetization has important consequences for domain wall dynamics and has to be taken into account.

We restrict our discussion to the longitudinal magnetoelectric response, in which the magnetization induced by the electric field is parallel to the AFM order parameter, irrespective of its spatial orientation. This is the case for the exchange-driven mechanism \cite{ME-review,Mostovoy,Mu-ME} of magnetoelectric response, which dominates in Cr$_2$O$_3$ and many other MEAFs at temperatures that are not too low. In Cr$_2$O$_3$ the only non-zero component of the magnetoelectric tensor in this approximation is $\alpha_\parallel=\alpha_{zz}$, where $z$ lies along the rhombohedral axis.\cite{Mostovoy,Mu-ME} It is assumed that the electric field is applied across an epitaxially grown (0001) film.

Adding the Berry-phase and magnetoelectric terms to the AFM Lagrangian,\cite{Bar,Haldane,Kim}, we can write the Lagrangian density of a MEAF, valid at low energies, as
\begin{equation}
\mathcal{L} = 2\epsilon\mathcal{J}\mathbf{a}(\mathbf{n}) \cdot \mathbf{\dot n}
	+ \frac12\left(\rho |\mathbf{\dot n}|^2 -A|\nabla \mathbf{n}|^2-\mathcal{K}_{\alpha\beta} n_\alpha n_\beta\right)
	- 2\epsilon\mathcal{J}\gamma \mathbf{H} \cdot \mathbf{n},\label{lag}
\end{equation}
where $\mathbf{n}$ is the unit vector in the direction of the AFM order parameter (staggered magnetization) $\mathbf L = (\mathbf M_1 - \mathbf M_2)/2$, $\mathbf{M}_1$ and $\mathbf{M}_2$ the sublattice magnetizations, $\mathcal{J}=L/(2\gamma)$ the angular momentum density on one sublattice, $\rho$ the effective inertia density, $A$ the exchange stiffness, and $\mathcal{K}_{\alpha\beta}$ the magnetocrystalline anisotropy tensor.\cite{K-note} Unless noted otherwise, it is assumed that the only nonzero component of this tensor is $\mathcal{K}_{zz}=-\mathcal{K}<0$. In the first term, $\epsilon=(M_1-M_2)/(M_1+M_2)=\alpha_\parallel E/L$, and $\mathbf{a(\mathbf{n})}$ is the vector potential of a magnetic monopole, $\nabla_{\mathbf{n}}\times\mathbf{a}=\mathbf{n}$; this term is the Berry-phase contribution from the small longitudinal magnetization $\mathbf{M}=(\mathbf{M}_1+\mathbf{M}_2)/2$ induced by the electric field.\cite{Kim,Altland} The last term in Eq.\ (\ref{lag}) is the magnetoelectric energy density;\cite{LL} $\gamma$ is the gyromagnetic ratio.

The AFM field theory at $E=0$ has characteristic scales of time, length, and pressure:
\begin{equation}
t_0=\sqrt{\rho/\mathcal{K}},\quad \lambda_0=\sqrt{A/\mathcal{K}},\quad \epsilon_0=\sqrt{A\mathcal{K}},
\end{equation}
which have direct physical meaning. $\epsilon_0$ is the scale of the domain wall energy per unit area. The magnon dispersion $\omega(k)=\sqrt{\omega_0^2+s^2k^2}$ has a gap $\omega_0=1/t_0$ and velocity $s=\lambda_0/t_0$. In Cr$_2$O$_3$, $\omega_0=0.68$ meV, \cite{Samuelsen} hence $t_0\approx1$ ps. The magnon velocity is $s=12$ km/s. \cite{Samuelsen} The length parameter $\lambda_0=st_0$ sets the scale of the domain wall width $d$. In Cr$_2$O$_3$ we find $\lambda_0=12$ nm and $d=\pi\lambda_0\approx 38$ nm.

The effective Lagrangian for low-energy domain wall dynamics is obtained by inserting the domain wall profile
\begin{equation}
\cos\theta(x)=\tanh \frac{x-X}{\lambda_0},\quad \phi(x)=\Phi,
\label{profile}
\end{equation}
parameterized by the collective variables $X$ and $\Phi$, in Eq.\ (\ref{lag}) and taking the integral over all space.
For the MEAF domain wall this leads to
\begin{equation}
L=\frac12 M\dot X^2+\frac12 I\dot\Phi^2+G\dot X \Phi - V(X,\Phi),
\end{equation}
where $M=2\rho/\lambda_0$ and $I=2\rho\lambda_0$ are the mass and moment of inertia per unit area of the wall, $V$ is the potential energy of the wall, which in a uniaxial AFM has no dependence of $\Phi$, and $G=4\epsilon\mathcal{J}$ is the gyrotropic term coupling the motion of the wall to its precession, which is proportional to $E$.

The equations of motion for the collective coordinates are
\begin{align}
M \ddot{X} &= - G \dot{\Phi} - \Gamma_{XX} \dot{X} + F,
\nonumber\\
I \ddot{\Phi} &= G \dot{X} - \Gamma_{\Phi\Phi} \dot{\Phi} + \tau,
\label{eom}
\end{align}
where $\Gamma_{XX}=4 \alpha_0 \mathcal J/\lambda_0$ and $\Gamma_{\Phi\Phi}=4 \alpha_0 \mathcal J \lambda_0$ are the viscous drag coefficients proportional to the Gilbert damping parameter $\alpha_0$, and $F=-\partial V/\partial X=2\alpha_{\parallel}E_zH_z=2\epsilon LH_z$. The torque $\tau = -\partial V/\partial\Phi$ vanishes in the case of uniaxial anisotropy. We will first consider the case $\tau = 0$ and then address the role of broken axial symmetry.

At $G=0$ we have a conventional AFM domain wall, which behaves as a massive particle subject to viscous drag, and whose angular collective variable $\Phi$ is completely passive.\cite{Kim}
However, the gyrotropic coupling $G$ induced by the electric field generates precession of the moving domain wall, which generates additional dissipation. In the steady state the moving domain wall precesses with the angular frequency
$\Omega = G \dot{X}/\Gamma_{\Phi\Phi}$, and the linear velocity of the wall is
\begin{equation}
v = \frac{F}{\Gamma_{XX} + G^2/\Gamma_{\Phi\Phi}}.\label{dotx}
\end{equation}
Thus, the additional dissipation induced by the gyrotropic coupling reduces the terminal velocity of the domain wall by the factor
$1 + G^2(\Gamma_{XX} \Gamma_{\Phi\Phi})^{-1}$.

Substituting the expressions for $\Gamma_{XX}$, $\Gamma_{\Phi\Phi}$, and $G$ in Eq.\ (\ref{dotx}), we obtain
\begin{equation}
v = \frac{2\epsilon/\alpha_0}{1 + (\epsilon/\alpha_0)^2}v_\mathrm{max},
\end{equation}
where $v_\mathrm{max}=\gamma H_z\lambda_0/2$. The maximum velocity $v_\mathrm{max}$ of the domain wall is reached at the optimal electric field strength $E_\mathrm{max}$ corresponding to $\epsilon=\alpha_0$. Interestingly, $v_\mathrm{max}$ depends neither on the magnetoelectric coefficient nor on the Gilbert damping constant.

Using the value $\gamma=1.76\times10^7$ s$^{-1}$/G and a reasonable field $H_z=100$ Oe, we find $v_\mathrm{max}\approx 10.6$ m/s. Assuming the switchable bit size of 50 nm, we estimate the switching time of about 5 ns. Note that the maximal MEAF domain wall mobility $v_\mathrm{max}/H_z\approx0.1$ m/(s$\times$Oe) is 2--3 orders of magnitude smaller in this regime compared to ferromagnets like permalloy.\cite{Konishi}

The Gilbert damping constant can be determined from the relation $T=\rho/(2\alpha_0\mathcal{J})$, where $T$ is the relaxation time.\cite{Kim} To estimate $T$ in Cr$_2$O$_3$, we use the width of the AFM resonance $\Delta H=900$ Oe,\cite{Foner} which translates into $\Delta\omega=1.6\times10^{10}$ s$^{-1}$ and $T=1/\Delta\omega\approx 60$ ps. Using the value $K=2\times10^5$ erg/cm$^3$,\cite{Foner} we find the inertia density $\rho=2Kt_0^2\approx 4\times10^{-19}$ g/cm. The value of $\mathcal{J}$ is obtained from the local magnetic moment\cite{Corliss} 2.76 $\mu_B$ and volume $\Omega\approx 50$ \AA$^3$ per formula unit. Putting these estimates together, we obtain $\alpha_0\approx2\times10^{-4}$.

The relation $\epsilon=\alpha_0$ then gives $E_\mathrm{max}\approx60$ V/$\mu$m in Cr$_2$O$_3$, where we used the peak value $\alpha_\parallel\approx10^{-4}$, which is reached at 260 K.
The magnetoelectric pressure corresponding to $E=E_\mathrm{max}$ and $H_z=100$ Oe is $F_\mathrm{max}=2\alpha_0 L H_z\approx 40$ erg/cm$^3$. To put this value in perspective, we note that in ferromagnetic iron the magnetic field of 100 Oe exerts a pressure of about $3\times10^5$ erg/cm$^3$ on the domain walls. The ``loss'' of four orders of magnitude in MEAF is due to the small magnitude of the magnetic moment induced by the electric field. Alternatively, one can say that a 100 Oe coercivity in a MEAF at $E\sim E_\mathrm{max}$ is equivalent, assuming similar material quality, to a 10 mOe coercivity in iron. Thus, it is clear that reasonably fast switching of a MEAF with uniaxial anisotropy requires samples of very high quality, unless the temperature is close to the N\'eel point $T_N$ where the domain wall width diverges and the coercivity becomes small even in low-quality samples. Indeed, isothermal MEAF switching has so far been observed only close to $T_N$.\cite{He}

In the presence of lattice imperfections, switching is possible if the magnetoelectric pressure $F$ applied to the domain wall exceeds the depinning pressure $F_c$. Since $T_N=307$ K of Cr$_2$O$_3$ is too low for passively-cooled computer applications, it needs to be either doped or strained to increase its $T_N$. In particular, boron doping on the Cr sublattice has been shown to raise $T_N$ significantly. \cite{Mu,Street} Random substitutional disorder in a doped material leads to an intrinsic pinning potential and nonzero coercivity. Let us estimate the effective depinning pressure for this representative case.

For simplicity, we assume that B dopants modify the exchange interaction locally but do not strongly affect the magnetocrystalline anisotropy. According to Ref.\ \onlinecite{Mu}, boron doping enhances the exchange coupling for the Cr atoms that have a B neighbor by a factor of 2--3. The concentration of B atoms is $n=3x/\Omega$, where $x$ is the B-for-O substitution concentration. Therefore, we make a crude estimate that the exchange stiffness $A$ is enhanced by a factor of 2 in regions of volume $2\Omega$, whose concentration is $n$.

Let $a^\ast$ be the radius of a sphere with volume $2\Omega$. The force acting on the domain wall from the vicinity of one B atom is $f\sim(a^\ast/\lambda_0)^3A$. The typical pinning force on a portion of the domain wall of size $R^2$ then becomes $f_\mathrm{pin}\sim \sqrt{n\lambda_0R^2f^2}$. The typical correlation length for the domain wall bending displacement is the Larkin length $R_c$, \cite{Larkin,Blatter,Chauve,Nattermann} which is found by equating $f_\mathrm{pin}$ to the typical elastic force $f_\mathrm{el}\sim u\sqrt{A\mathcal{K}}$ produced by the domain wall, where $u\sim\lambda_0$ corresponds to the situation in which the domain wall deforms weakly. This gives $R_c\sim \sqrt{\frac{\lambda_0 A\mathcal{K}}{nf^2}}$. The depinning threshold can then be estimated as $F_c\sim A/R_c^2=n\lambda_0 A (2\Omega/\lambda_0^3)^2$. Using $x=0.03$ and $A\sim 10^{-6}$ erg/cm, we find $F_c\sim 10$ erg/cm$^3$, which is comparable to the magnetoelectric pressure at $H=100$ Oe and $E=E_\mathrm{max}$, as we have estimated above. Other imperfections may further increase $F_c$. Thus, as expected from the comparison with typical ferromagnets, even weak pinning associated with homogeneous doping can impede MEAF switching. This sensitivity to lattice disorder, along with the low upper bound on the domain wall mobility, present serious challenges for the implementation of magnetoelectric devices.

We will now show that both of these limitations can be overcome by introducing a relatively small in-plane anisotropy component $\mathcal{K}_{yy}=\mathcal{K}_\perp$ in addition to the axial component $\mathcal{K}_{zz}=-\mathcal{K}$. Such in-plane anisotropy can be induced by applying a small in-plane shear strain to the magnetoelectric crystal, for example, by using a piezoelectric element, an anisotropic substrate, or anisotropic thermal expansion in a patterned structure. The physics of domain wall motion at $\mathcal{K}_\perp\neq0$ is similar to Walker breakdown in ferromagnets, where the anisotropy with respect to $\Phi$ appears due to the magnetostatic interaction. \cite{Walker}

In the equations of motion (\ref{eom}) we now have, after integrating out the domain wall profile (\ref{profile}), a nonzero torque $\tau=-\lambda_0 \mathcal{K}_\perp\sin2\Phi$ per unit area. There is a steady-state solution with $\dot\Phi=0$ and $v=F/\Gamma_{XX}$, as long as $v<v_W$, where $v_W/v_\mathrm{max}=2(\mathcal{K}_\perp/F_\mathrm{max})^{1/2}$ is analogous to the Walker breakdown velocity.\cite{Walker}
For example, in order to achieve $v_W\sim100$ m/s, we need to have $\mathcal{K}_\perp\gtrsim900$ erg/cm$^3$, which is three orders of magnitude smaller than $\mathcal{K}$. It is likely that $\mathcal{K}_\perp$ of this order can be achieved with a fairly small in-plane shear strain.

\begin{figure}[htb]
\includegraphics[width=0.75\columnwidth]{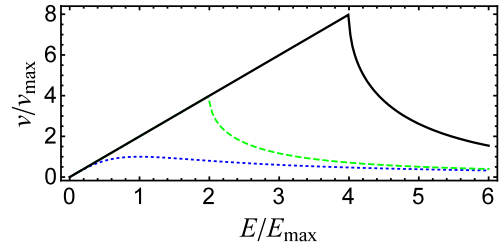}
\caption{Average domain wall velocity $v$ as a function of $E/E_\mathrm{max}$ at $\mathcal{K}_\perp=0$ (dotted blue line), $4F_\mathrm{max}$ (dashed green), and $16F_\mathrm{max}$ (solid).}
\label{vvsf}
\end{figure}

Below the Walker breakdown the domain wall velocity is linear in $E$: $v/v_\mathrm{max}=2E/E_\mathrm{max}$. At $F>\Gamma_{XX}v_W$ the in-plane anisotropy can no longer suppress domain wall precession, so that its velocity becomes oscillatory. The average velocity has a cusp at $F=\Gamma_{XX}v_W$ and declines with a further increase in $F$, as shown in Fig.\ \ref{vvsf}.

In the presence of $\mathcal{K}_\perp\gtrsim 900$ erg/cm$^3$, the fields $E\approx 0.2$ V/nm and $H\approx100$ Oe result in $v\approx 70$ m/s and $F\approx140$ erg/cm$^3$. Under these conditions the switching time of a nanoscale bit can be well below a nanosecond, while the magnetoelectric pressure $F$ exceeds the intrinsic depinning field of B-doped Cr$_2$O$_3$ by an order of magnitude. Clearly, the imposition of in-plane anisotropy offers compelling advantages for device applications by improving switchability and speed.

It is interesting to note that the domain wall mobility can be changed by orders of magnitude by imposing a non-zero $K_\perp$ in the strong-electric-field regime $\epsilon\gg\alpha_0$. This peculiar feature of MEAF domain wall dynamics can be directly checked experimentally.

Devices based on MEAF switching offer a distinct advantage in terms of energy efficiency. Energy dissipated when a bit is switched is $E_\mathrm{dis}=2 \alpha_{\parallel} E_z H_z V =FV$, where $V$ is the switched volume. This is the energy difference between the two AFM domain states of the bit. Taking the switching volume to be a cube with a 50 nm edge, we estimate $E_\mathrm{dis}\sim 10^{-14}$ erg for the field magnitudes chosen above. This corresponds to an upper limit on the intrinsic power consumption of 1 mW/Gbit, assuming that each bit is switched every nanosecond. Clearly, energy dissipation in a magnetoelectric memory device would be dominated by losses in the external circuitry.

\begin{figure}[htb]
\includegraphics[width=0.92\columnwidth]{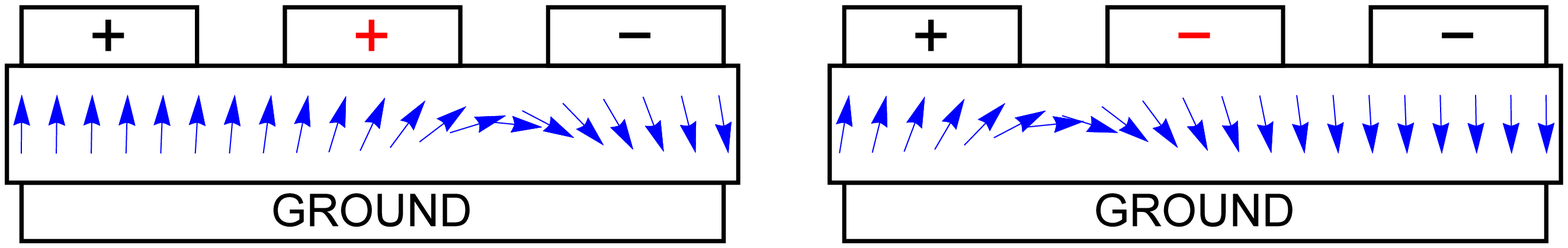}
\caption{Two logical states of a split-gate magnetoelectric memory bit. Central region: MEAF layer. Arrows: AFM order parameter $\mathbf{L}$. The continuous bottom electrode is grounded. The left and right set gates labeled ``+'' and ``--'' are activated during the write operation, enforcing fixed AFM domain states underneath these gates. The state of the bit is recorded by the voltage applied to the central control gate. The permanent magnetic field is applied vertically (not shown).}
\label{split}
\end{figure}

As we argued above, fast memory operation should be based on domain wall-mediated switching. Therefore, it is necessary to design the architecture of a bit in such a way that the domain wall is not annihilated at the surface as the bit is switched. One way to achieve this is through the use of a multiple-gate scheme, as shown in Fig.\ \ref{split}. In this scheme, additional ``set gates'' are used to initialize and maintain two different AFM states at the edges of the active magnetoelectric layer, which are labeled $+$ and $-$ in Fig.\ \ref{split}, thereby trapping the domain wall inside the device. The set gates need to be activated only during the write operation, along with the control gate. Positive or negative voltage applied to the control gate selects the AFM domain state in the switched area and drives the domain wall between the positions shown by the dashed lines. This scheme is somewhat reminiscent of the spin-transfer torque domain wall device.\cite{Currivan} The control gate can also provide the memory read function by employing a FM layer, coupled, via the boundary magnetization of the MEAF, to its AFM domain state, and a spin valve, or a similar magnetoresistive element, grown on top of it. Alternatively, the AFM domain state can be detected through the anomalous Hall effect in a thin non-magnetic control gate.\cite{Kosub}

Since the domain wall should fit inside the bit, its width $d$ sets a limitation for the downward scaling of the length of the MEAF element. The width of this element, however, can be significantly smaller. To facilitate downscaling, the domain wall width $d$ can be reduced by increasing the magnetocrystalline anisotropy of the MEAF. For example, it is known that the addition of Al increases $K$ in Cr$_2$O$_3$.\cite{Al-Cr2O3}

To increase the memory density, the basic element shown in Fig.\ \ref{split} may be assembled in a linear array, for example by using a sequence of gates like $+$C$-$C$+$C$\dots$ where $+$ and $-$ denote the set gates and C the control gates. In this way each internal set gate protects the domain walls on both sides, and for a long array the footprint reduces from 3 to 2 gates per bit. Alternatively, the use of several control gates in sequence allows for more than two positions for each domain wall and leads to memory density $(\log_2 n)/n$ bits per gate, where $n$ is the number of control gates in a sequence. The memory density is lowest for $n=3$ but the gain compared to $n=2$ or $n=4$ is only about 6\%. If all gates are made identical, a linear array offers an additional possibility for reprogramming, i.e., for designating different gates as $+$ and $-$ set gates; this could be implemented by applying sufficiently long voltage pulses to the new set gates to allow reliable switching. Using the bottom electrode, or sections of it, for magnetic readout could also allow for additional majority-gate functionality. Thus, a multiple split-gate architecture could provide combined memory and logic capabilities.

To conclude, we have described the domain wall dynamics in a MEAF and discussed its implications for magnetoelectric memory applications. We found that the domain wall mobility $v/H$ in a uniaxial MEAF reaches a maximum at a certain electric field $E_\mathrm{max}$ and then declines, which is unfavorable for device applications. However, the domain wall mobility and switchability can be greatly improved by imposing a small in-plane anisotropy, which blocks the domain wall precession, and using electric fields $E\sim0.2$ V/nm. A split-gate architecture is proposed to trap the domain wall inside the bit element. A linear gate array extending this architecture can offer advantages in memory density, programmability, and logic functionality integrated with nonvolatile memory. While the domain-wall-driven mechanism allows reliable and fast switching, it limits the minimum length of the bit to the domain wall width.

We thank Christian Binek and Peter Dowben for useful discussions and KITP for hospitality. KB acknowledges support from the Nanoelectronics Research Corporation (NERC), a wholly-owned subsidiary of the Semiconductor Research Corporation (SRC), through the Center for Nanoferroic Devices (CNFD), an SRC-NRI Nanoelectronics Research Initiative Center, under Task ID 2398.001. KB and AK acknowledge support from the NSF through the Nebraska Materials Research Science and Engineering Center (MRSEC) (Grant No.\ DMR-1420645). AK acknowledges support from the DOE Early Career Award {DE-SC0014189}. OT acknowledges support from the DOE Office of Basic Energy Sciences, Division of Materials Sciences and Engineering under Award DE-FG02-08ER46544. OAT acknowledges support by the Grants-in-Aid for Scientific Research (Nos.\ 25800184, 25247056 and 15H01009) from the MEXT, Japan and SpinNet. This research was also supported in part by NSF under Grant No.\ PHY11-25915.


\begin{thebibliography}{99}
\bibitem{K-note} $\hat{\mathcal{K}}$ is defined as in Ref.\ \onlinecite{Kim} for consistency; it is related to the conventional $\hat K$ as $\hat{\mathcal{K}}=2\hat K$.
\end{thebibliography}
\end{document}